\documentclass[prd,nofootinbib]{revtex4}

\usepackage{amsmath}
\usepackage{graphicx}

\newcommand{\tr}{\mathop{\rm tr}\nolimits}
\newcommand{\const}{\mathop{\rm const}\nolimits}
\newcommand{\Pexp}{\mathop{\rm Pexp}\nolimits}
\newcommand{\inst}{\mathop{\rm inst}\nolimits}
\newcommand{\T}{\mathop{\rm T}\nolimits}
\newcommand{\NP}{\mathop{\rm NP}\nolimits}
\newcommand{\PP}{\mathop{\rm P}\nolimits}
\newcommand{\eff}{\mathop{\rm eff}\nolimits}
\newcommand{\dia}{\mathop{\rm dia}\nolimits}
\newcommand{\para}{\mathop{\rm para}\nolimits}
\newcommand{\cl}{\mathop{\rm cl}\nolimits}
\newcommand{\cool}{\mathop{\rm cool}\nolimits}

\newcommand{\sing}{\mathop{\rm sing}\nolimits}

\newcommand{\I}{\mathop{\rm I}\nolimits}

\begin{document}

\begin{flushright}
ITEP--PH--5/2001
\end{flushright}
\vspace{1.5cm}

\title{Instanton IR stabilization in the nonperturbative confining vacuum}
\author{N.O.~Agasian}
\email{agasyan@heron.itep.ru}
\author{S.M.~Fedorov}
\email{fedorov@heron.itep.ru}
\affiliation{
Institute of Theoretical and Experimental Physics,\\
117218, Moscow, B.Cheremushkinskaya 25, Russia}
\date{December 14, 2001}

\begin{abstract}

The influence of nonperturbative fields on instantons in
quantum chromodynamics is studied. Nonperturbative vacuum is described in
terms of nonlocal gauge invariant vacuum averages of gluon field strength.
Effective action for instanton is derived in bilocal approximation and
it is demonstrated that stochastic background gluon fields are responsible for infra-red (IR) stabilization of
instantons. Dependence of characteristic instanton size on gluon condensate and correlation
length in nonperturbative vacuum is found. It is shown that instanton size in QCD is of order
of 0.25~fm. Comparison of obtained instanton size distribution with lattice data is made.
\end{abstract}

\maketitle

\section{INTRODUCTION}
\label{sec_intr}
Instanton is the first explicit example of nonperturbative (NP) quantum
fluctuation of gluon field in QCD. It was introduced in 1975 by Polyakov and coauthors~\cite{BPST}.
An important development originated with 't~Hooft's classic paper~\cite{tHooft_76}, in which he calculated
the semi-classical tunneling rate.
Instanton gas as a model of QCD vacuum was proposed in pioneer works by Callan, Dashen and Gross~\cite{CDG_76,CDG_78}.
These topologically nontrivial field configurations
are essential for the solution of some problems of quantum chromodynamics.
Instantons allow to explain anomalous breaking of $U(1)_A$ symmetry and the $\eta'$ mass~\cite{tHooft_76b,Witten_Venez_79}.
Spontaneous chiral symmetry breaking (SCSB) can be explained with the help of instanton and anti-instanton field
configurations in QCD vacuum~\cite{Diak_Pet_86}. An important role of instantons in scalar and pseudoscalar
channels was pointed out in paper~\cite{GI}. Taking into account instantons is of crucial importance for many
phenomena of QCD (see~\cite{Scha_Shur_98} and references therein).

At the same time, there is a number of serious problems in instanton physics. The first is the divergence of
integrals over instanton size $\rho$ at big $\rho$. This makes it impossible to calculate instantons'
contribution to some physical quantities, such as vacuum gluon condensate. Second, ''area law'' for Wilson loop
can not be explained in instanton gas model, hence quasiclassical instanton anti-instanton vacuum
lacks confinement which is responsible for hadron spectra.

There were many theoretical works aimed to solve the problem of instanton instability in it's size $\rho$.
To some extent they all were based on an attempt to stabilize instanton ensemble due
to effects of interaction between pseudoparticles. The most popular is the model of ''instanton liquid'',
which was phenomenologically formulated by Shuryak~\cite{Shuryak_81}. It states
that average distance between pseudoparticles is $\bar{R}\sim 1$~fm and their average
size is $\bar{\rho}\sim 1/3$~fm. Thus, $\bar{\rho}/\bar{R}\simeq 1/3$ and vacuum consists of
well separated, and therefore not very much deformed, instantons and anti-instantons. Quantitatively close
results were obtained by Diakonov and Petrov~\cite{Diak_Pet_84}. In their approach
stabilization is related to discovered by them classical repulsion between pseudoparticles. However,
further development~\cite{BY} revealed that instanton ensemble can not be stabilized
due to purely classical interaction.Thus, the mechanism for the suppression of
large-size instantons in the ensemble of topologically non-trivial fields is still
not understood.

On the other hand, QCD vacuum contains not only quasiclassical
instantons, but also other nonperturbative fields. The
investigation of instantons' behavior in stochastic QCD vacuum,
which is parametrized by ''vacuum correlators'' (i.e. nonlocal
gauge invariant averages of gluon field strength) was started in
Refs.~\cite{Agas_Sim_95, Agasian_96}. It was shown, that in nonperturbative
vacuum standard perturbation theory for instanton changes , which
leads to ''freezing'' of effective coupling
constant. It was then found in~\cite{Agasian_96} that nonlocal
interaction between large-size instanton ($\rho > 1$~fm)
and nonperturbative stochastic background field does not lead to
it's infrared inflation.

The aim of this work is to demonstrate that instanton can be stabilized in
nonperturbative vacuum and exist as a stable topologically nontrivial
field configuration against the background of stochastic nonperturbative fields,
which are responsible for confinement, and to find quantitatively it's size.
We evaluate nonlocal effective action for instanton in NP vacuum in gauge invariant way,
describing vacuum in terms of ''vacuum correlators'', and show that
in bilocal approximation effective action has a minimum at the definite instanton size $\rho_c$ .
The value of $\rho_c$ is in functional dependence on properties of bilocal correlator
$\langle G(x)\Phi(x,y)G(y)\rangle$, i.e. it depends on two parameters:
$\langle G^2\rangle$ -- gluon condensate and it's ''measure of inhomogeneity ''
$T_g$ -- correlation length in the condensate.

\section{GENERAL FORMALISM}
\label{sec_general}
The influence of NP fluctuations on instanton can be separated into two parts.
First, perturbation theory undergoes a change in NP background, which results in the
change of standard one-loop renormalization of instanton action. Second,
a direct nonlocal interaction of instanton with NP background field exists.

Standard euclidian action of gluodynamics has the form
\begin{equation}
\label{eq_action}
S[A]=\frac{1}{2g_0^2} \int d^4 x \tr(F_{\mu\nu}^2[A])=
\frac{1}{4} \int d^4 x F_{\mu\nu}^a[A]F_{\mu\nu}^a[A],
\end{equation}
where $ F_{\mu\nu}[A]=\partial_{\mu}A_{\nu} - \partial_{\nu}A_{\mu}-i[A_{\mu},A_{\nu}]$
is the strength of gluon field and we use the Hermitian matrix form for gauge fields
$A_\mu(x)= g_0A^a_\mu(x) {t^a}/{2}$ and $\tr t^at^b=\delta^{ab}/2$. We decompose $A_{\mu}$ as
\begin{equation}
\label{eq_decomp}
A_{\mu} = A_{\mu}^{\inst}+B_{\mu}+a_{\mu},
\end{equation}
where $A_{\mu}^{\inst}$ is an instanton-like field configuration with a unit topological charge
$Q_{\T}[A^{\inst}]=1$; $a_{\mu}$ is quantum field (expansion in $a_{\mu}$ reduces to
perturbation theory, which in gluodynamics leads to asymptotic freedom);
$B_{\mu}$ is nonperturbative background field (with zero topological charge), which can be
parametrized by gauge invariant nonlocal vacuum averages of gluon field
strength\footnote{In operator product expansion method and in QCD sum rules nonperturbative
field is characterized by a set of local gluon condensates $\langle G^2\rangle$,
$\langle G^3\rangle$, \ldots}.

In general case effective action for instanton in NP vacuum takes the form
\begin{equation}
\label{eq_genrl}
Z=e^{-S_{\eff}[A^{\inst}]} = \int
[Da_{\mu}]\left\langle e^{-S[A^{\inst}+B+a]}\right\rangle,
\end{equation}
where $\langle...\rangle$ implies averaging over background field $B_{\mu}$,
\begin{equation}
\label{eq_measure}
\left\langle\hat{O}(B)\right\rangle=\int d\mu(B) \hat{O}(B)
\end{equation}
and $d\mu(B)$ is the measure of integration over NP fields, explicit form of which is not
important for the following consideration.

Expanding $S[A]$ up to the second power in $a_{\mu}$ one obtains
$Z=\langle Z_1(B)Z_2(B)\rangle$, where
\begin{align}
&Z_1(B)=e^{-S[A^{\inst}]} \int
  [Da_{\mu}] \det (\bar{\nabla}^2_{\mu}) \exp\left[\frac{1}{g^2_0}\tr\int d^4 x
  \left\{-(\bar{\nabla}_{\mu} a_{\nu})^2+2i\bar{F}_{\mu\nu}[a_{\mu},a_{\nu}]\right\}\right]
\label{eq_z1_z2_a}\\
&Z_2(B)=\exp\{-S[A^{\inst}+B]+S[A^{\inst}]\}
\label{eq_z1_z2_b}
\end{align}
Here we use the notation $\bar{A}\equiv A^{\inst}+B$, $\bar{F}_{\mu\nu}\equiv F_{\mu\nu}[\bar{A}]$,
and $\bar{\nabla}_{\mu}a_{\nu}=\partial_{\mu} a_{\nu}-i[\bar{A}_{\mu},a_{\nu}]$ is a covariant
derivative.
Integration over $a$ and $B$ in~(\ref{eq_z1_z2_a}),(\ref{eq_z1_z2_b}) corresponds to
averaging over fields that are responsible for the physics at different scales.
Integration over $a_{\mu}$ takes into account perturbative gluons and
describes phenomena at small distances. Averaging over $B_{\mu}$ (formally
interaction with gluon condensate) accounts for
phenomena at scales of confinement radius. Therefore it is physically clear that averaging
factorizes and one obtains
\begin{equation}
\label{eq_factoriz}
Z \to \langle Z_1(B)\rangle \langle Z_2(B)\rangle
\end{equation}
It should be noted that in the limit of infinite number of colors $N_c \to \infty$
factorization~(\ref{eq_factoriz}) is exact
$Z(N_c \to \infty) = \langle Z_1(B)\rangle \langle Z_2(B)\rangle$.
This allows us to write effective action of instanton in NP vacuum as a sum
of two terms, ''perturbative'' and ''nonperturbative'':
\begin{align}
\label{eq_seff_gnrl_a}
&S_{\eff}[A^{\inst}]=S^{\PP}_{\eff}[A^{\inst}]+S^{\NP}_{\eff}[A^{\inst}]\\
\label{eq_seff_gnrl_b}
&S^{\PP}_{\eff}[A^{\inst}]=-\ln\langle Z_1(B)\rangle\\
\label{eq_seff_gnrl_c}
&S^{\NP}_{\eff}[A^{\inst}]=-\ln\langle Z_2(B)\rangle=
-\ln\left\langle \exp \{-S[A^{\inst}+B]+S[A^{\inst}]\}\right\rangle
\end{align}

\section{INTERACTION BETWEEN INSTANTON AND NONPERTURBATIVE
VACUUM GLUON FIELDS}
\label{sec_IRstab}

The general expression for
one-instanton field configuration has the well known form
\begin{equation}
\label{eq_inst}
A^{\inst}_{\mu} = 2 t^{b} R^{b \beta}
\overline{\eta}^{\beta}_{\mu\nu}
\frac{(x-x_0)_{\nu}}{(x-x_0)^2} f\left(\frac{(x-x_0)^{2}}{\rho^2}\right),
\end{equation}
where matrix $R^{b \beta}$ ensures embedding of instanton into $SU(N_c)$ group,
$b = 1,2,\,.\,.\,.N_c^2-1$; $\beta = 1,2,3$,~~
$\overline{\eta}^{\alpha}_{\mu\nu}$ are 't~Hooft symbols. $R^{b\beta}$ satisfied relations
$f^{abc}R^{b\beta}R^{c\gamma} =\varepsilon^{\beta\gamma\delta}R^{a\delta}$,~~
$R^{b\beta}R^{b\gamma}=\delta^{\beta\gamma}$ and $f^{abc}R^{a\alpha}R^{b\beta}R^{c\gamma} =
\varepsilon^{\alpha\beta\gamma}$.
In singular gauge profile function $f(z)$ satisfies boundary conditions $f(0)=1$, $f(\infty)=0$ and
the classical solution has the form
\begin{equation}
f(z)=\frac{1}{1+z^2}
\label{eq_profile}
\end{equation}
The probability to find an instanton is determined by the
classical action functional $S_{\cl}[A]$ for the solution~(\ref{eq_inst}),(\ref{eq_profile}); that is
\begin{equation}
w\sim\exp\{-S_{\cl} [A^{\inst}_\mu]\}=\exp\{-8\pi^2/g^2_0\}
\end{equation}
The preexponential factor was calculated in~\cite{tHooft_76}. The result for one-instanton
contribution to the QCD partition function is $Z_{\I}= \int  dn(\rho,x_0,R)$,
where $dn$ is the differential instanton density
\begin{align}
&dn(\rho,x_0,R)=[dR] d^4 x_0\frac{d\rho}{\rho^5} d_0(\rho)\\
&d_0(\rho)=\frac{4.6\exp\{-1.68 N_c\}}{\pi^2(N_c-1)!(N_c-2)!}
\left(\frac{8\pi^2}{g^2(\rho)}\right)^{2N_c}\exp \left
\{-\frac{8\pi^2}{g^2(\rho)}\right \}
\label{eq_d0}
\end{align}
In the two-loop approximation in gluodynamics, the coupling
constant $g^2(\rho)$ is given by
\begin{equation}
\frac{8\pi^2}{g^2(\rho)}=
b\ln \left(\frac{1}{\rho\Lambda}\right) +\frac{b_1}{b}
\ln\ln\left(\frac{1}{\rho\Lambda}\right)+ O(1/\ln(1/\rho\Lambda)),
~~~b=\frac{11}{3}N_c,~~b_1=\frac{17}{3} N_c
\label{eq_coupling}
\end{equation}
$\Lambda\equiv\Lambda_{PV}$ corresponds to the Pauli-Villars regularization
scheme, $\Lambda\sim 200$~MeV.

Callan, Dashen and Gross~\cite{CDG_78} were first to show that in a constant gauge
field instanton behaves as a colored four-dimensional dipole. Shifman, Vainshtein and
Zakharov~\cite{SVZ} generalized this result to the case of
interaction between a small-size instanton, $\rho<
0.2$~fm, and nonperturbative long-wave fluctuations that
are described by the local vacuum condensate $\langle G^2\rangle$. As a
result, $d_0 (\rho)$ is replaced by $d_{\eff}(\rho)$, where
\begin{equation}
d_{\eff} (\rho)\propto (\Lambda\rho)^{b}\left(1+\frac{4\pi^4\langle
G^2\rangle}{(N^2_c-1)g^4} \rho^4+...\right)
\label{eq_d0_mod}
\end{equation}
Thus, we arrive at the well-known problem of the infrared inflation of
instantons in $\rho$.

It was shown in~\cite{Agas_Sim_95, Agasian_96} that  in NP vacuum standard perturbation theory
for instantons changes, which results in ''freezing'' of effective coupling constant. The perturbative
part of effective instanton action in stochastic vacuum $S^{\PP}_{\eff}[A^{\inst}]$
was shown to be
\begin{equation}
\label{eq_seff_pt_final}
S^{\PP}_{\eff}(\rho)=\frac{b}{2} \ln\frac{1/\rho^2+m_*^2}{\Lambda^2}
\end{equation}
Here $m_*\simeq 0.75 m_{0^{++}}  \sim 1\mbox{GeV}$, where $0^{++}$ is the lightest glueball.
It follows from~(\ref{eq_seff_pt_final})
that for small-size instantons a standard perturbative result
$S_{\eff}^{\PP}(\rho\ll 1/m_{*})={8 \pi^2}/{g^2(\rho)}$ is recovered, and for large-size instantons
$S^{\PP}_{\eff}(\rho>1/m_{*})\to \const$. Correspondingly, factor $\sim (\Lambda\rho)^{b}$ in~(\ref{eq_d0_mod})
ceases to grow with the increase of $\rho$. Next, large-size instanton
$\rho>\rho_0=(192/\langle G^2\rangle)^{1/4}\sim 1$~fm was considered
in~\cite{Agasian_96}. It's field is weak compared to characteristic field strengths in
gluon condensate $\langle G^2\rangle$, and it was considered as a perturbation in~\cite{Agasian_96}.
It was demonstrated that due to instanton's nonlocal interaction with gluon condensate
there is no infrared inflation for $\rho>\rho_0$.

We consider effect of NP fields on instanton not assuming that it's field is weak (for
arbitrary $\rho$), i.e. we evaluate $\langle Z_2\rangle$.
In this work we make use of the method of vacuum correlators,
introduced in works of Dosch and Simonov~\cite{Dosch_87}. NP vacuum of QCD is described in terms of
gauge invariant vacuum averages of gluon fields (correlators)
$$
\Delta_{\mu_1 \nu_1 ... \mu_n \nu_n}=\langle\tr
G_{\mu_1\nu_1}(x_1) \Phi(x_1,x_2) G_{\mu_2\nu_2}(x_2) ...
G_{\mu_n\nu_n}(x_n) \Phi(x_n,x_1)\rangle,
$$
where $G_{\mu\nu}$ is gluon field strength, and $\Phi(x,y)=\Pexp\left(i\int\limits_y^x B_{\mu}dz_{\mu}\right)$
is a parallel transporter, which ensures gauge invariance. In many cases bilocal approximation appears
to be sufficient for qualitative and quantitative description of various physical phenomena in QCD.
Moreover, there are indications that corrections due to higher correlators are small and amount to
several percent~\cite{DShS}.

The most general form of bilocal correlator, which follows from
antisymmetry in Lorentz indices, is given by
\begin{align}
\label{eq_bilocal}
&\langle g^2 G_{\mu\nu}^a(x,x_0) G_{\rho\sigma}^b(y,x_0)\rangle
= \langle G^2 \rangle \frac{\delta^{ab}}{N_c^2-1} \times \notag\\
&\quad \times\left\{
\frac{D(z)}{12}(\delta_{\mu\rho}\delta_{\nu\sigma}-\delta_{\mu\sigma}\delta_{\nu\rho})
+\frac{\overline{D}(z)}{6}(n_{\mu}n_{\rho}\delta_{\nu\sigma}+n_{\nu}n_{\sigma}\delta_{\mu\rho}-
n_{\mu}n_{\sigma}\delta_{\nu\rho}-n_{\nu}n_{\rho}\delta_{\mu\sigma})\right\},
\end{align}
where $G_{\mu\nu}(x,x_0)=\Phi(x_0,x)G_{\mu\nu}(x)\Phi(x,x_0)$,~~
$n_{\mu}={z_{\mu}}/{|z|}={(x-y)_{\mu}}/{|x-y|}$ is the unit vector,
$\langle G^2 \rangle \equiv \langle g^2 G_{\mu\nu}^a G_{\mu\nu}^a \rangle$
and, as it follows from normalization, $D(0)+\overline{D}(0)=1$.

Functions $D(z)$ and $\overline{D}(z)$ have both perturbative and
nonperturbative contributions. We will consider only
nonperturbative part, because perturbative one has already been
taken into account in $S^{\PP}_{\eff}$. Most data about
nonperturbative components of these functions come from numerical
simulations on lattice. Gluon condensate $\langle G^2 \rangle$ is
also determined from lattice calculations, but there exists a
widely used estimate for this value based on charmonium spectrum
analysis and QCD sum rules $\langle G^2 \rangle \simeq 0.5
\mbox{GeV}^4$~\cite{SVZ_79}. According to the lattice data $D(z)$
and $\overline{D}(z)$ are exponentially decreasing functions
$D(z)=A_0 \exp(-z/T_g)$, $\overline{D}(z)=A_1 z \exp(-z/T_g)/T_g$,
where $T_g$ is the gluonic correlation length, which was measured
on the lattice~\cite{DiG} and estimated analytically~\cite{23} to
be $T_g\sim 0.2$~fm. Besides, according to lattice measurements
$A_1 \ll A_0$ ($A_1 \sim A_0/10$). Lattice data from paper of
Di~Giacomo~\cite{DiGiacomo_2000} are presented in
Table~\ref{tab_digiacomo}. $SU(3)$~full stands for chromodynamics
with 4 quarks, while $SU(2)$ and $SU(3)$~quenched mean pure
$SU(2)$ and $SU(3)$ gluodynamics, respectively. Note, that
$\langle G^2\rangle = 0.87$~GeV$^{4}$ for $SU(3)$~full is in good
agreement with more recent calculations based on QCD sum
rules~\cite{NN}.

\begin{table}[!htb]
\caption{Lattice data~\cite{DiGiacomo_2000} for bilocal correlator}
\label{tab_digiacomo}
\begin{tabular}{lcc}
\hline\hline
& $\langle G^2\rangle$, GeV$^4$ & $T_g$,~fm\\
\hline
$SU(2)$ quenched & 13 & 0.16 \\
$SU(3)$ quenched & 5.92 & 0.22 \\
$SU(3)$ full & 0.87 & 0.34\\
\hline\hline
\end{tabular}
\end{table}

To evaluate $S_{\eff}^{\NP}$ defined in~(\ref{eq_seff_gnrl_c}) we use the cluster expansion,
which is well known in statistical physics~\cite{IH}:
\begin{equation}
\label{eq_cluster}
\langle \exp(x) \rangle =
 \exp \left(\sum_n \frac{\langle\!\langle x^n \rangle\!\rangle}{n!} \right),
\end{equation}
where $\langle x \rangle = \langle\!\langle x \rangle\!\rangle$;~~~
$\langle x^2 \rangle = \langle\!\langle x^2 \rangle\!\rangle +
  \langle x \rangle^2$;~~~
$\langle x^3 \rangle = \langle\!\langle x^3 \rangle\!\rangle
  +3\langle x \rangle \langle\!\langle x^2 \rangle\!\rangle +
  \langle x \rangle^3$;~~~\ldots

Next, we modify expression~(\ref{eq_seff_gnrl_c}) for $S_{\eff}^{\NP}$
by adding $S[B]$ to the exponent:
\begin{equation}
S^{\NP}_{\eff}[A^{\inst}]=
-\ln\left\langle \exp \{-S[A^{\inst}+B]+S[A^{\inst}]+S[B]\}\right\rangle
\end{equation}
In bilocal approximation this is equivalent to a change of normalization of
partition function\footnote{Indeed, starting from~(\ref{eq_seff_gnrl_c}) one finds that in bilocal
approximation~(\ref{eq_seff_4}) gets additional terms of the type $\langle G^2 \rangle \int d^4 x$,
which do not depend on $A^{\inst}$ and therefore result only in renormalization of $Z_2$.},
which is not important for the following consideration.

Nonperturbative part of the effective instanton action $S_{\eff}^{\NP}$ takes the form
\begin{align}
\label{eq_seff_3}
&S_{\eff}^{\NP}=\langle S[A^{\inst}+B]-S[B]-S[A^{\inst}]\rangle - \notag\\
&\frac{1}{2} \Bigl( \left\langle
(S[A^{\inst}+B]-S[B]-S[A^{\inst}])^2 \right\rangle -\left\langle
S[A^{\inst}+B]-S[B]-S[A^{\inst}] \right\rangle^2 \Bigr) + \ldots
\end{align}

By using Fock-Shwinger gauge $x_{\mu}A_{\mu}=x_{\mu}B_{\mu}=0$ (instanton field $A_{\mu}^{\inst}$
satisfies it owing to the  properties of 't~Hooft symbols)
and taking into account that
\begin{align}
\label{eq_Sab-sb}
S[A^{\inst}+B]-S[B]-S[A^{\inst}] =& \frac{1}{2g^2} \int d^4 x \tr \Bigl\{
- ([A^{\inst}_{\mu},B_{\nu}]-[A^{\inst}_{\nu},B_{\mu}])^2 \notag\\
&+2 F_{\mu\nu}[A^{\inst}]G_{\mu\nu}[B] - 4 i
(F_{\mu\nu}[A^{\inst}]+G_{\mu\nu}[B])[A^{\inst}_{\mu},B_{\nu}]
\Bigr\}
\end{align}
one gets
\begin{equation}
\label{eq_seff_4}
S_{\eff}^{\NP}=S_{\dia}+\frac{1}{2}S_{\dia}^2+S_{\para}+S_1+S_2,
\end{equation}
where in bilocal approximation
\begin{align}
&S_{\dia}=-\frac{1}{2g^2}\int d^4 x
\left\langle\tr\left[\left([A_{\mu},B_{\nu}]-[A_{\nu},B_{\mu}]\right)^2\right]\right\rangle \label{eq_sdia}\\
&S_{\para}=-\frac{1}{2g^4}\int d^4 x d^4 y \left\langle
\tr\left(F_{\mu\nu}(x)G_{\mu\nu}(x)\right)
\tr\left(F_{\rho\sigma}(y)G_{\rho\sigma}(y)\right) \right\rangle \label{eq_spara}\\
&S_{1}=\frac{2}{g^4}\int d^4 x d^4 y \left\langle \tr\left(
F_{\mu\nu}[A_{\mu},B_{\nu}]\right)_x
\tr\left(
F_{\rho\sigma}[A_{\rho},B_{\sigma}]
\right)_y \right\rangle  \label{eq_s1}\\
&S_{2}=\frac{2i}{g^4}\int d^4 x d^4 y\left\langle
\tr\left(F_{\mu\nu}G_{\mu\nu}\right)_x
\tr\left(F_{\rho\sigma}[A_{\rho},B_{\sigma}]\right)_y
\right\rangle \label{eq_s2}
\end{align}
We use notations $S_{\dia}$ (diamagnetic) and $S_{\para}$ (paramagnetic)
for contributions~(\ref{eq_sdia}) and~(\ref{eq_spara}) into interaction
of instanton with background field. Physical motivation for this is discussed in
detail in~\cite{Agasian_96}. For the sake of simplicity we drop further the index
'inst' for instanton field, $A_{\mu}^{\inst}\equiv A_{\mu}$.

In Fock-Shwinger gauge $B_{\mu}(x)=x_{\nu}\int\limits_0^1 \alpha d  \alpha G_{\nu\mu}(\alpha x)$,
and that allows us to substitute vacuum averages in~(\ref{eq_sdia})-(\ref{eq_s2}) with correlators.

The general form of the instanton field is
\begin{equation}
\label{eq_inst_1}
A_{\mu}(x)=\Phi(x,x_0)A^{\sing}_{\mu}(x-x_0)\Phi(x_0,x),
\end{equation}
where $\Phi(x,y)=\Pexp\left(i\int\limits_y^x B_{\mu}dz_{\mu}\right)$ is a parallel transporter,
and $A^{\sing}_{\mu}(x-x_0)$ is instanton field in a fixed (singular) gauge~(\ref{eq_inst}),(\ref{eq_profile}).
Inserting~(\ref{eq_inst_1}) into~(\ref{eq_sdia})-(\ref{eq_s2}) and using representation~(\ref{eq_bilocal})
for gauge invariant condensate $\langle g^2 G_{\mu\nu}^a(x,x_0) G_{\rho\sigma}^b(y,x_0)\rangle$
one gets
\begin{align}
\label{eq_seff_component_2_a}
&S_{\dia} = \frac{\langle G^2 \rangle}{12} \frac{N_c}{N_c^2-1} \int
d^4 x \int\limits_0^1 \alpha d \alpha \int\limits_0^1 \beta d
\beta \, x^2 (A_{\mu}^a(x))^2 [D((\alpha-\beta)x)+2\overline{D}((\alpha-\beta)x)] \\
\label{eq_seff_component_2_b}
&S_{\para}=-\frac{\langle G^2 \rangle}{48 g^2} \frac{1}{N_c^2-1}
\int d^4 x d^4 y \left[F_{\mu\nu}^a(x) D(x-y) F_{\mu\nu}^a(y) +
4\frac{(x-y)_{\mu}(x-y)_{\rho}}{(x-y)^2}F_{\mu\nu}^a(x)\overline{D}(x-y)F_{\rho\nu}^a(y)\right]\\
\label{eq_seff_component_2_c}
&\begin{array}{rl}
  S_{1}=&-\frac{\langle G^2 \rangle}{24}
  \frac{1}{N_c^2-1} \int d^4 x d^4 y \int\limits_0^1
  \alpha d \alpha \int\limits_0^1 \beta d \beta \,(xy
  \delta_{\nu\sigma}-x_{\sigma}y_{\nu})
  f^{abe}f^{cde}\\
  &\times F_{\mu\nu}^a(x)A_{\mu}^b(x)F_{\rho\sigma}^{c}(y)
  A_{\rho}^{d}(y)D(\alpha x - \beta y) + O(\overline{D})
\end{array}\\
\label{eq_seff_component_2_d}
&S_{2}=\frac{\langle G^2\rangle}{12 g}\frac{1}{N_c^2-1} \int d^4 x
d^4 y \,
f^{abc}F_{\rho\sigma}^a(y)A_{\rho}^b(y)F_{\sigma\nu}^c(x)y_{\nu}\int\limits_0^1
\alpha d \alpha D(x-\alpha y)+ O(\overline{D})
\end{align}

Thus we have obtained effective action for instanton in NP vacuum in bilocal
approximation\footnote{The tensor structure of instanton was used to
derive~(\ref{eq_seff_component_2_a})-(\ref{eq_seff_component_2_d}).}. It can be seen
from numerical calculation that typical instanton size in QCD is $\rho_c\sim 0.25$~fm.
For such $\rho$ instanton field in it's center is strong
$F_{\mu\nu}^2(x=x_0, \rho_c)=192/\rho_c^4\gg\langle G^2\rangle$, and therefore classical
instanton solution is not much deformed in the area $|x|<\rho$, which gives the main
contribution to integrals~(\ref{eq_seff_component_2_a})-(\ref{eq_seff_component_2_d}).

The problem of asymptotic behavior of instanton solution far from the center $|x|\gg\rho$ was
studied in detail in Refs.~\cite{Diak_Pet_84, MAK, Dor}. Our numerical analysis shows that
the value of $\rho_c$ is almost not affected by the asymptotic of classical instanton solution
provided that condensate $\langle G^2\rangle$ and correlation length $T_g$ have reasonable values.

Next we have to define function $D(z)$. We use Gaussian form
$D(z)=\exp({-\mu^2 z^2})$, $\mu\equiv{1}/{T_g}$. $D(z)$ is a monotonously decreasing function with
typical correlation length $T_g$, and numerical calculations~\cite{AF} show that explicit
form of $D(z)$ is not important. For instance, substitution of $\exp(-\mu^2 z^2)$ with $\exp(-\mu z)$
almost does not affect the value of $\rho_c$.

We use the standard profile function $f={\rho^2}/(\rho^2+x^2)$. Certainly,
exact instanton profile (i.e. profile that minimizes action) is different from this
one. Nevertheless, knowing the dependence of $S_{\eff}$ on $\rho$ we will be able
to make a conclusion about instanton behavior (inflation or stabilization) and
to determine it's typical size in NP vacuum.

Thus we obtain after integration over space directions
\begin{align}
\label{eq_component_4_a}
&S_{\dia}=\frac{2\pi^2}{g^2} \frac{\langle
G^2\rangle}{\mu^4}\frac{N_c}{N_c^2-1} \zeta^4
\int\limits_0^{\infty}dx
\frac{x^3}{(x^2+\zeta^2)^2} \varphi(x) \\
\label{eq_component_4_b}
&S_{\para}=\frac{-16\pi^4}{g^4}\frac{\langle
G^2\rangle}{\mu^4}\frac{1}{N_c^2-1} \zeta^4\int\limits_0^{\infty}
dx dy \frac{x^2 y^2
e^{-(x^2+y^2)}I_3(2xy)}{(x^2+\zeta^2)^2(y^2+\zeta^2)^2}\\
&S_1=\frac{-64\pi^4}{g^4} \frac{\langle
  G^2\rangle}{\mu^4}\frac{1}{N_c^2-1}
  \zeta^8\int\limits_0^{\infty} dx dy \frac{x^2
  y^2}{(x^2+\zeta^2)^3(y^2+\zeta^2)^3}
   \int\limits_0^1 d\alpha \int\limits_0^1 d\beta
  e^{-(\alpha^2 x^2 + \beta^2 y^2)}(I_1(2\alpha\beta
  xy)+I_3(2\alpha\beta xy))\\
&S_2=\frac{-64\pi^4}{g^4} \frac{\langle
  G^2\rangle}{\mu^4}\frac{1}{N_c^2-1} \zeta^6\int\limits_0^{\infty} dx dy \frac{x^2
  y^2 e^{-x^2}}{(x^2+\zeta^2)^2(y^2+\zeta^2)^3}\int\limits_0^1 d\alpha
  e^{-\alpha^2 y^2} I_3(2\alpha xy)
\end{align}
where $I_n(x)=e^{-\frac{i\pi n}{2}}J_n(i x)$ is Bessel function of imaginary argument
(Infeld function), $\zeta \equiv \mu \rho$  and
\begin{equation}
\label{eq_varphi}
\varphi(x)=\int\limits_0^1 \alpha d \alpha \int\limits_0^1 \beta
d \beta e^{-(\alpha-\beta)^2 x^2} =
e^{-x^2}\left(\frac{1}{3x^2}-\frac{1}{6x^4}\right) +
\frac{1}{6x^4} - \frac{1}{2x^2} + \frac{2}{3x} \Phi(x)
\end{equation}
where $\Phi(x)=\int\limits_0^x e^{-\xi^2}d\xi$ is error function.

We studied asymptotic behavior of instanton effective action and it's dependence on dimensionless parameter
$\zeta \equiv \mu \rho$. For a small-size instanton $\rho\ll 1/\mu$ ($\zeta \rightarrow 0$) we found
for $S_{\eff}^{\NP}$
\begin{align}
&S_{\dia}\to -\frac{\pi^2}{2 g^2}
\frac{N_c}{N_c^2-1}\langle G^2\rangle \rho^4 \ln(\mu\rho),\\
&S_{\para}+S_1 +S_2\to -\frac{2 \pi^4}{g^4}
\frac{\langle G^2\rangle}{N_c^2-1} \rho^4 +O\left(\langle G^2\rangle
\rho^4 (\mu \rho)\right).
\end{align}

In the opposite limit of large $\rho\gg 1/\mu$ ($\zeta \rightarrow \infty$) one has

\begin{align}
&S_{\dia}\to \frac{\pi^{7/2}}{6 g^2}
\frac{N_c}{N_c^2-1}\langle G^2\rangle  \frac{\rho^3}{\mu},\\
&S_{\para}+S_1 +S_2\to \const.
\end{align}

Differential instanton density is proportional to  $\frac{dn} {d^4 z d\rho} \propto \exp(-S_{\eff})$.
Thus, ''diamagnetic'' interaction of instanton with NP fields leads to
its infrared stabilization in size $\rho$.

\section{DISCUSSION AND CONCLUSION}
\label{DISCUSSION_CONCLUSION}

\begin{figure}[!htb]
\begin{picture}(288,188)
\put(0,10){\includegraphics{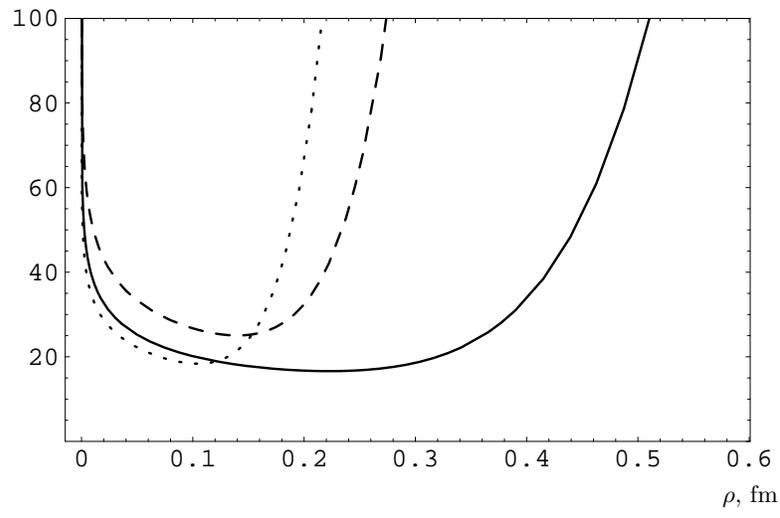}}
\put(270,0){$\rho$, fm}
\end{picture}
\caption{Effective action $S_{\eff}(\rho)$.\\
$SU(3)$ full (solid line), $SU(2)$ quenched (dotted line)  and $SU(3)$ quenched (dashed line)}
\label{fig_1}
\end{figure}

\begin{figure}[!htb]
\begin{picture}(288,188)
\put(0,10){\includegraphics{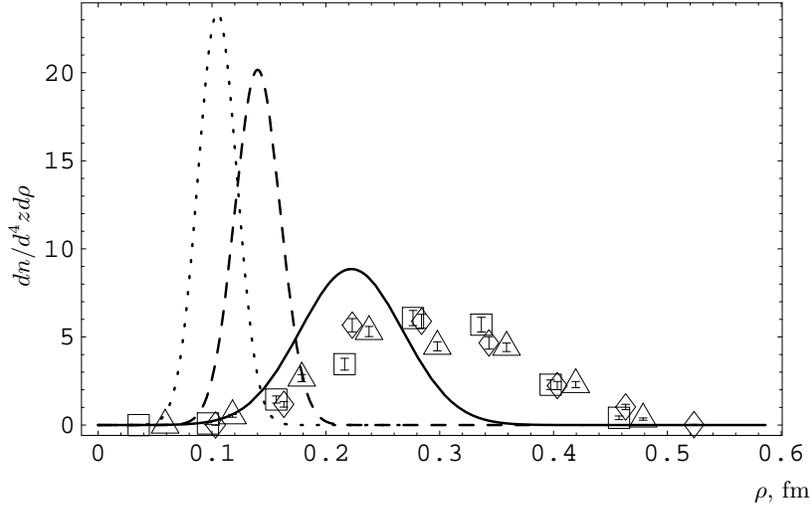}}
\put(270,0){$\rho$, fm}
\put(-12,80){\rotatebox{90}{$dn/d^4 z d\rho$}}
\end{picture}
\caption{Instanton density $dn/d^4 z d\rho$ and lattice data~\cite{Hasen_Niet_98}.\\
$SU(3)$ full (solid line), $SU(2)$ quenched (dotted line)  and $SU(3)$ quenched (dashed line)}
\label{fig_2}
\end{figure}

In this work we have studied instanton behavior in NP vacuum, which was parametrized
with gauge invariant vacuum averages of gluon field strength. We have derived effective instanton
action in bilocal approximation and have shown that ''diamagnetic'' term $S_{\dia}$ leads to IR
stabilization of instanton. Numerical results for $S_{\eff}$ are plotted in Fig.~\ref{fig_1}.
Three curves correspond to $SU(2)$ and $SU(3)$ gluodynamics and $SU(3)$ QCD with quarks.
Instanton size distribution $dn/d^4 z d\rho \sim \exp(-S_{\eff})$ and corresponding lattice
data~\cite{Hasen_Niet_98} are presented in Fig.~\ref{fig_2}. All graphs are normalized to
the commonly accepted instanton density 1~fm$^{-4}$.
It should be mentioned that over the last years the lattice results have not
much converged (compare the recent conference~\cite{Latt_conf}).
Different groups roughly agree on instantons size within a factor
of two, e.g. $\bar{\rho}=0.3 \ldots 0.6$~fm for $SU(3)$ gluodynamics.
There is no agreement at all concerning the density $N/V$.
As a tendency, lattice studies give higher density and larger
instantons than phenomenologically assumed.

\begin{figure}[!htb]
\begin{picture}(288,188)
\put(0,10){\includegraphics{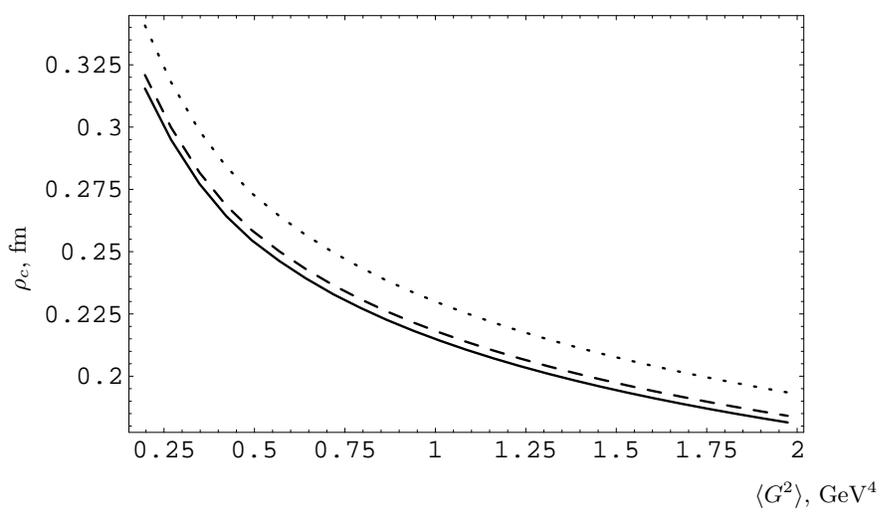}}
\put(270,0){$\langle G^2 \rangle$, GeV$^4$}
\put(-12,80){\rotatebox{90}{$\rho_c$, fm}}
\end{picture}
\caption{Instanton size as a function of gluon condensate ($N_c=3$, $N_f=4$) at
$T_g=0.2$~fm (dotted line), $T_g=0.3$~fm (dashed line), $T_g=0.34$~fm (solid line)}
\label{fig_3}
\end{figure}

\begin{figure}[!htb]
\begin{picture}(288,188)
\put(0,10){\includegraphics{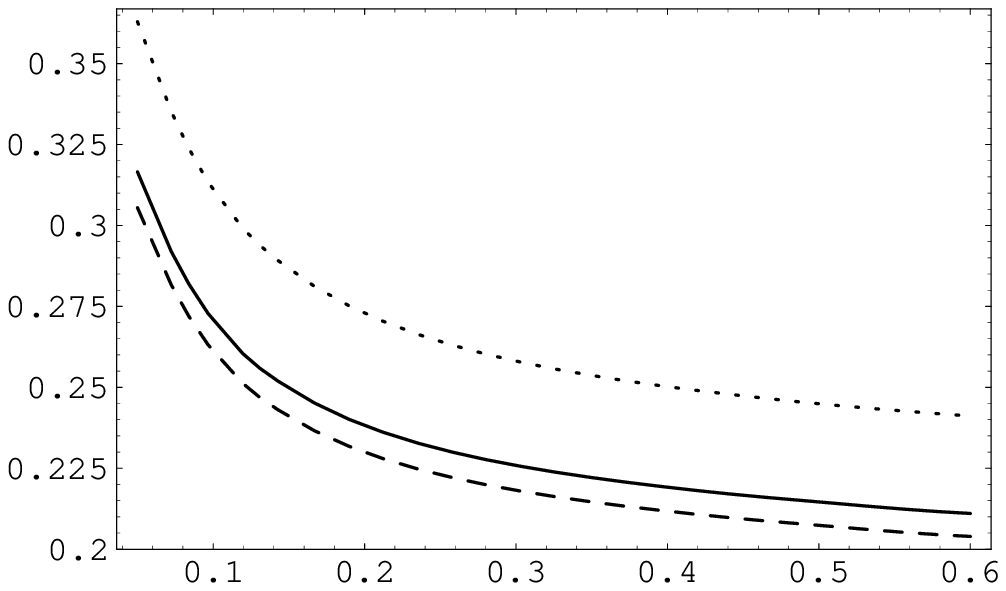}}
\put(270,0){$T_g$, fm}
\put(-12,80){\rotatebox{90}{$\rho_c$, fm}}
\end{picture}
\caption{Instanton size as a function of correlation length ($N_c=3$, $N_f=4$) at
$\langle G^2 \rangle=0.5$~GeV$^4$ (dotted line), $\langle G^2 \rangle=0.87$~GeV$^4$ (solid line),
$\langle G^2 \rangle=1.0$~GeV$^4$ (dashed line)}
\label{fig_4}
\end{figure}

Using our model we find for $\langle G^2\rangle = 5.92$~GeV$^4$ and $T_g=0.22$~fm
that $\rho_c \simeq 0.15$~fm, which is less than phenomenological ($\sim 0.3$~fm)
and lattice results. However, we can present physical arguments to
explain these deviations. Lattice calculations include cooling procedure,
during which some lattice configurations of gluon field are discarded.
This procedure can result in a change in gluon condensate $\langle G^2\rangle$,
and thus instanton size distribution is calculated at a value of gluon condensate
$\langle G^2\rangle_{\cool}$ which is smaller than physical value $\langle G^2\rangle$
\footnote{Besides, cooling probably affects correlation length $T_g$ as well.}.
Therefore, lattice data for average instanton size $\bar{\rho}$ should
be compared with our calculations for $\rho_c$, performed at
smaller values of $\langle G^2\rangle$. We show dependence of $\rho_c$ on
$\langle G^2\rangle$ for several values of $T_g$ in Fig.~\ref{fig_3}.
One can see that increase of $\langle G^2\rangle$ results in
decrease of instanton size, and that effect is a result of nonlocal
''diamagnetic'' interaction of instanton with NP fields. Fig.~\ref{fig_4}
shows $\rho_c$ as a function of $T_g$ for several values of $\langle G^2\rangle$.
The smaller $T_g$, the larger instanton is. It is physically clear that
less correlated NP fields ($T_g \to 0$) have smaller influence on instanton
field configuration, which occupies 4D euclidian volume with characteristic size $\rho$
($\rho \gg T_g$).
On the other hand, perturbative quantum fluctuations tend to inflate instanton,
and therefore $\rho_c$ increases with the decrease of $T_g$. In full
QCD (see Table~\ref{tab_digiacomo}) we find $\rho_c \simeq 0.25$~fm.

We did not go beyond bilocal approximation in this work. As mentioned above,
this approximation is good enough not only for qualitative, but also
for quantitative (with accuracy of several percent~\cite{DShS}) description
of some phenomena in nonperturbative QCD. In the problem under consideration
there are two small parameters. These are $1/g^2(\rho_c) \sim 0.15 \ldots 0.25$ and
$1/N_c$, and there powers grow in each term of cluster expansion. Moreover,
we made an estimate for the sum of leading terms in cluster expansion~\cite{AF},
and found that IR stabilization stays intact ($\rho_c$ appears to
be a little smaller). Thus, proposed model describes physics of
single instanton stabilization in NP vacuum, not only qualitatively,
but also quantitatively with rather good accuracy.

\section*{ACKNOWLEDGMENTS}

We are grateful to E.-M.~Ilgenfritz, A.B.~Kaidalov,
V.A.~Novikov and Yu.A.~Simonov for useful discussions and comments.
The financial support of RFFI grant 00-02-17836 and
INTAS grant CALL 2000 N 110 is gratefully acknowledged.

\newpage

\begin{thebibliography}{99}

\bibitem{BPST}
A.~M.~Polyakov,
Phys.\ Lett.\ {\bf B59} (1975) 79;~
%
A.~A.~Belavin, A.~M.~Polyakov, A.~S.~Shvarts and Y.~S.~Tyupkin,
Phys.\ Lett.\ {\bf B59} (1975) 85.

\bibitem{tHooft_76}
G.~'t~Hooft,
Phys.\ Rev.\ {\bf D14} (1976) 3432.

\bibitem{CDG_76}
C.~G.~Callan, R.~F.~Dashen and D.~J.~Gross,
Phys.\ Lett.\ {\bf B63} (1976) 334.

\bibitem{CDG_78}
C.~G.~Callan, R.~F.~Dashen and D.~J.~Gross,
Phys.\ Rev.\ {\bf D17} (1978) 2717;~
%
Phys.\ Rev.\ {\bf D19} (1979) 1826.


\bibitem{tHooft_76b}
G.~'t~Hooft,
Phys.\ Rev.\ Lett.\  {\bf 37} (1976) 8.

\bibitem{Witten_Venez_79}
E.~Witten,
Nucl.\ Phys.\ {\bf B149} (1979) 285;~
%
G.~Veneziano,
Nucl.\ Phys.\ {\bf B159} (1979) 213.

\bibitem{Diak_Pet_86}
D.~Diakonov and V.~Y.~Petrov,
Nucl.\ Phys.\ {\bf B272} (1986) 457.

\bibitem{GI}
B.~V.~Geshkenbein and B.~L.~Ioffe,
Nucl.\ Phys.\ {\bf B166} (1980) 340.

\bibitem{Scha_Shur_98}
T.~Schafer and E.~V.~Shuryak,
Rev.\ Mod.\ Phys.\  {\bf 70} (1998) 323
[hep-ph/9610451].

\bibitem{Shuryak_81}
E.~V.~Shuryak,
Nucl.\ Phys.\ {\bf B203} (1982) 93.

\bibitem{Diak_Pet_84}
D.~Diakonov and V.~Y.~Petrov,
Nucl.\ Phys.\ {\bf B245} (1984) 259.

\bibitem{BY}
I.~I.~Balitsky and A.~V.~Yung,
Phys.\ Lett.\ {\bf B168} (1986) 113.

\bibitem{Agas_Sim_95}
N.~O.~Agasian and Yu.~A.~Simonov,
Mod.\ Phys.\ Lett.\ {\bf A10} (1995) 1755.

\bibitem{Agasian_96}
N.~O.~Agasian,
Phys.\ Atom.\ Nucl.\  {\bf 59} (1996) 297.

\bibitem{SVZ}
M.~A.~Shifman, A.~I.~Vainshtein and V.~I.~Zakharov,
Nucl.\ Phys.\ {\bf B163} (1980) 46.

\bibitem{Dosch_87}
H.~G.~Dosch,
Phys.\ Lett.\ {\bf B190} (1987) 177;~
%
H.~G.~Dosch and Yu.~A.~Simonov,
Phys.\ Lett.\ {\bf B205} (1988) 339;~
%
Yu.~A.~Simonov,
Nucl.\ Phys.\ {\bf B307} (1988) 512.

\bibitem{DShS}
A.~Di~Giacomo, H.~G.~Dosch, V.~I.~Shevchenko and Yu.~A.~Simonov,
hep-ph/0007223.


\bibitem{SVZ_79}
M.~A.~Shifman, A.~I.~Vainshtein and V.~I.~Zakharov,
Nucl.\ Phys.\ {\bf B147} (1979) 385,~448.

\bibitem{DiG}
M.~Campostrini, A.~Di~Giacomo and G.~Mussardo,
Z.\ Phys.\ {\bf C25} (1984) 173;~
%
A.~Di~Giacomo and H.~Panagopoulos,
Phys.\ Lett.\ {\bf B285} (1992) 133;~
%
A.~Di~Giacomo, E.~Meggiolaro and H.~Panagopoulos,
hep-lat/9603017;~
%
M.~D'Elia, A.~Di~Giacomo and E.~Meggiolaro,
Phys.\ Lett.\ {\bf B408} (1997) 315
[hep-lat/9705032].

\bibitem{23}
Yu.~A.~Simonov,
Few Body Syst.\  {\bf 25} (1998) 45
[hep-ph/9712248];~
%
Yu.~A.~Simonov,
Phys.\ Atom.\ Nucl.\  {\bf 61} (1998) 855.

\bibitem{DiGiacomo_2000}
A.~Di~Giacomo,
hep-lat/0012013.

\bibitem{NN}
S.~Narison,
Phys.\ Lett.\ {\bf B387} (1996) 162
[hep-ph/9512348].

\bibitem{IH}
Kerzon Huang, Statistical Mechanics, Jon Wiley and Sons, Inc., New York--London (1963);~
%
A.~Isihara, Statistical Physics, Academic Press, New York--London (1971).

\bibitem{MAK}
A.~B.~Migdal, N.~O.~Agasian and S.~B.~Khokhlachev,
JETP Lett.\  {\bf 41} (1985) 497;~
%
N.~O.~Agasian and S.~B.~Khokhlachev,
Sov.\ J.\ Nucl.\ Phys.\  {\bf 55} (1992) 628,~633;~
%
N.~O.~Agasian,
hep-ph/9803252;~
%
hep-ph/9904227.

\bibitem{Dor}
A.~E.~Dorokhov, S.~V.~Esaibegian, A.~E.~Maximov and S.~V.~Mikhailov,
Eur.\ Phys.\ J.\ {\bf C13} (2000) 331
[hep-ph/9903450].

\bibitem{AF}
N.O.~Agasian and S.M.~Fedorov, ITEP-PH-6/2001

\bibitem{Hasen_Niet_98}
A.~Hasenfratz and C.~Nieter,
Phys.\ Lett.\ {\bf B439} (1998) 366 [hep-lat/9806026].

\bibitem{Latt_conf}
J.~W.~Negele,
Nucl.\ Phys.\ Proc.\ Suppl.\  {\bf 73} (1999) 92
[hep-lat/9810053];~
%
M.~Teper,
Nucl.\ Phys.\ Proc.\ Suppl.\  {\bf 83} (2000) 146
[hep-lat/9909124];~
%
M.~Garcia Perez,
Nucl.\ Phys.\ Proc.\ Suppl.\  {\bf 94} (2001) 27
[hep-lat/0011026].
\end{thebibliography}
\end{document}